\documentclass[aps,floatfix]{revtex4}
\usepackage{amsmath,amssymb}
\usepackage{epsfig}

\newcommand{\bra}[1]{\langle #1 |}
\newcommand{\ket}[1]{| #1 \rangle}

\begin{document}
\title{Matrix rearrangement approach for the entangling power with mixed qudit systems}
\author{Xiao-Ming Lu, Xiaoguang Wang, Yang Yang, and Jian Chen}
\affiliation{Institute of Modern Physics, Department of Physics, Zhejiang University, Hangzhou 310027, China}
\begin{abstract}
We extend the former matrix rearrangement approach of the entangling power to the general cases, without the requirement of the same dimensions of the subsystems. The entangling power of a unitary operator is completely determined by its realignment and partial transposition.  As applications, we calculate the entangling power for the Ising interaction and the isotropic Heisenberg interaction in the mixed qudit system.

\end{abstract}
\maketitle

\section{Introduction}
Entanglement is one of the most important quantum correlations among subsystems. It plays a key role in quantum information processing and is considered as a kind of resource for quantum computation and communication~\cite{Nielsen}. In recent decades, many efforts are being put into the quantification, generation and enhancement of the entanglement.

From the aspect of generation of entanglement, what we concern is that given an operation, what is its capability of producing entanglement. Recently, the entanglement capability of quantum evolution and Hamiltonians is studied in Refs.~\cite{Dur,Kraus,Cirac,Hammerer,Vidal,Dur2,XWang,Makhlin}. Generally speaking, for unitary operators, we can let them act on pure product states, then observe the amount of the entanglement produced. It always depends on the input states. There have been two ways to obtain a input-state-independent quantity for measuring the unitary operators' capacity of entanglement creation. One is averaging over the product input state \cite{Cirac,Zanardi,XWang1,Lakshminarayan}, the other is taking the maximum of the entanglement produced over the product input states \cite{Leifer,Chefles}.

The entangling power belongs to the first kind. It is defined as \cite{Zanardi}
\begin{equation}
  \label{eq:ep}
  e_{\mathrm{p}}(U)\equiv \overline{E(U\ket{\psi_1}\otimes\ket{\psi_2})},
\end{equation}
where the average is over all the product states distributed according to some probability distribution. The linear entropy $E$ is adopted as entanglement measurement for pure states of bipartite systems hereafter, which is defined as
\begin{equation}
E(\ket{\Psi}) = 1-\mathrm{Tr}_1(\rho_1^2),
\end{equation}
where $\rho_1=\mathrm{Tr}_2(\ket{\Psi} \bra{\Psi})$ is the reduced density matrix of the first subsystem.

When we study the closed quantum systems, under Schr\"{o}dinger equation, the time evolution is unitary. Especially, when the Hamiltonian is time-independent, the unitary time evolution operator is in the one-parameter group generated by the Hamiltonian, i.e., $U(t)=\exp(-iHt)$. So we can also consider conveniently the time average of entangling power which is define as
\begin{equation}
  \label{eq:time_average_ep}
  \overline{e_{\mathrm{p}}}(U(t))=\lim_{T\rightarrow \infty}\frac{1}{T}\int\limits_0^T e_{\mathrm{p}}(U(t))dt.
\end{equation}
This entangling power not only average over all initial product states, but also over all time range.

Most recently, a matrix rearrangement approach to the entangling power was given in \cite{ZMa} for subsystems with equal dimensions. It expresses the entangling power in terms of the realigned matrix \cite{KChen} and the partially transposed matrix \cite{Peres}. These two matrix rearrangements are originally used to study the separability problem of the quantum mixed states, and lead to two useful criterion for entanglement \cite{Peres,Horodecki,Rudolph,KChen}.

On the other hand, because the entanglement plays an intriguing role in the kinematic process of the reducing of density operators~\cite{Popescu,Goldstein} and the reduced dynamics~\cite{Breuer1}, the entangling power has also been used to study some dynamical systems, especially in the chaotic systems~\cite{XWang2,Demkowicz}.

In this paper, first, we prove that the entangling power of an unitary operator $U$ is completely determined by its realignment and partial transposition, without requirement of same dimensions of the subsystems. Then we apply it to calculate the entangling power of the Ising interaction and the isotropic Heisenberg interaction in hybrid qudit $d_1 \times d_2$ systems. The analytic results are obtained. We can see the exact relation between the time-average entangling power and dimensions of the Hilbert spaces of the subsystems. At last, a conclusion is given.

\section{Matrix Rearrangement approach for the entangling power }
\noindent
After averaging with respect to the uniform distribution of the non-entangled pure initial states, for a unitary operator on $d_1\times d_2$ systems, the entangling power is given by \cite{Zanardi}
\begin{equation}
  \label{eq:original}
\begin{split}
   &e_{\mathrm{p}}(U)=1-C_{d_1}C_{d_2}\sum \limits_{\alpha = 0,1} I_{\alpha} (U),\\
   &I_\alpha(U)=\mathrm{Tr}(T_{1+ \alpha,3+ \alpha})+\langle U^{\otimes2},(T_{1+\alpha,3+\alpha})U^{\otimes 2}T_{13}\rangle,
\end{split}
\end{equation}
where $C_d^{-1}=d(d+1)$, and $ \langle A , B \rangle  \equiv \mathrm{Tr}(A^\dag B) $ is the Hilbert-Schmidt scalar product. Let $T_{i,j}(i,j=1,\ldots ,4)$ denote the transposition between the \textit{i}th and the \textit{j}th factor of $\mathcal{H}^{\otimes 2} \cong (C^{d_1} \otimes C^{d_2}) \otimes (C^{d_1} \otimes C^{d_2}) $. For example,
\begin{equation}
  T_{13} (\ket{i} \otimes \ket{j} \otimes \ket{k} \otimes \ket{l})= \ket{k} \otimes \ket{j} \otimes \ket{i} \otimes \ket{l}.
\end{equation}
It is easy to check that $\mathrm{Tr}(T_{13})=d_1d_2^2$ and $ \mathrm{Tr}(T_{24})=d_1^2d_2$.
Then we get another form of the entangling power
\begin{equation}
  \label{eq:general_ep}
  e_{\mathrm{p}}(U)=\mathcal{F}_{d_1,d_2}-\mathcal{G}_{d_1,d_2}\sum\limits_{\alpha=0,1}\langle U^{\otimes2},(T_{1+\alpha,3+\alpha})U^{\otimes 2}T_{13}\rangle,
\end{equation}
where
\begin{align}
  \nonumber
  \mathcal{F}_{d_1,d_2}&=\frac{d_1d_2+1}{(d_1+1)(d_2+1)},\\
  \mathcal{G}_{d_1,d_2}&=\frac{1}{d_1d_2(d_1+1)(d_2+1)}.
\end{align}

For the $d\times d$ systems ($d_1=d_2$), the entangling power can be expressed as \cite{Zanardi_2}
\begin{equation}
    e_{\mathrm{p}}=(\frac{d}{d+1})^2[E(U)+E(US_{12})-E(S_{12})],
\end{equation}
where $E(O)$ is the entanglement of the operator $O$ and can be expressed using the matrix realignment as $E(O)=1-\mathrm{Tr}[(U^R (U^R)^\dag)^2]/d^4$,
 so the entangling power has the following form \cite{ZMa}
\begin{equation}
  \label{eq:Ma}
  \begin{split}
    e_{\mathrm{p}}(U)=&\frac{d^2+1}{(d+1)^2}-\frac{1}{d^2(d+1)^2}\{\mathrm{Tr}[(U^R(U^R)^\dag)^2]+\mathrm{Tr}[(U^{T_1}(U^{T_1})^\dag)^2]\},
  \end{split}
\end{equation}
where $R$ is the matrix realignment \cite{KChen} and $T_1$ the partial transposition with respect to the first subsystem \cite{Peres}. They are defined by
\begin{equation}
  \begin{split}
    (U^R)_{ij,kl}&=(U)_{ik,jl},\\
    (U^{T_1})_{ij,kl}&=(U)_{kj,il},
  \end{split}
\end{equation}
respectively.

We now show that the entangling power is determined by the realigned and partially transposed unitary operator without requiring
the same dimensions $d_1=d_2$.
From Eq.~(\ref{eq:general_ep}), we see that the terms
related to the evolution operator are only $\langle U^{\otimes
2},T_{1+\alpha,3+\alpha}U^{\otimes 2}T_{13}\rangle$. So the task
left is to get the other expressions of them. A unitary operator can be written as
\begin{align}
  \label{eq:U}
  \nonumber
  U&=U_{i\alpha,j\beta}e_{ij}\otimes e_{\alpha \beta}\\
  \nonumber
  &=\Lambda _{ij,\alpha \beta}e_{ij}\otimes e_{\alpha \beta}\\
  &=\Gamma_{j\alpha,i\beta}e_{ij}\otimes e_{\alpha \beta},
\end{align}
where $e_{ij}=\ket{i} \bra{j}$. Repeated indices are summed in all
cases. Here we use English letter subfixes for the first
subsystem, and Greek letter ones for the second. Obviously, in the
expression above, $\Lambda = U^R$,and $\Gamma = U^{T_1}$.
Then we get
\begin{align}
  \label{eq:detail}
  \nonumber
  U^{\otimes 2}&=\Lambda_{ij,\alpha \beta}\Lambda_{kl,\gamma \delta}e_{ij}\otimes e_{\alpha \beta}\otimes e_{kl}\otimes e_{\gamma \delta}\\
  \nonumber
  &=\Gamma_{j \alpha,i \beta}\Gamma_{l \gamma,k \delta}e_{ij}\otimes e_{\alpha \beta}\otimes e_{kl}\otimes e_{\gamma \delta},\\
  \nonumber
  T_{13}U^{\otimes 2}T_{13}&=\Lambda_{ij,\alpha \beta}\Lambda_{kl,\gamma \delta}e_{kl}\otimes e_{\alpha \beta}\otimes e_{ij}\otimes e_{\gamma \delta},\\
  T_{24}U^{\otimes 2}T_{13}&=\Gamma_{j \alpha,i \beta}\Gamma_{l \gamma,k \delta}e_{il}\otimes e_{\gamma \beta}\otimes e_{kj}\otimes e_{\alpha \delta}.
\end{align}

Substituting Eq.~(\ref{eq:detail}) into the related terms, and notice the normalization relation $\langle e_{ij},e_{kl}\rangle = \delta _{ik} \delta _{jl}$, we get
\begin{equation}
    \langle U^{\otimes 2},T_{13}U^{\otimes 2}T_{13} \rangle = \Lambda_{ij,\alpha \beta}^* \Lambda_{kl,\gamma \delta}^* \Lambda_{kl,\alpha \beta} \Lambda_{ij,\gamma \delta} = \mathrm{Tr}[(\Lambda \Lambda^\dag)^2],
\end{equation}
and
\begin{equation}
    \langle U^{\otimes 2},T_{24}U^{\otimes 2}T_{13} \rangle = \Gamma_{j \alpha,i \beta}^* \Gamma_{l \gamma,k \delta}^* \Gamma_{l \gamma,i \beta} \Gamma_{j \alpha, k \delta}=\mathrm{Tr}[(\Gamma \Gamma^\dag)^2].
\end{equation}
Substituting them into Eq.~(\ref{eq:general_ep}), note that
$\Lambda = U^R$, and $\Gamma = U^{T_1}$, we get the general
expression of the entangling power
\begin{equation}
  \label{eq:ep_form}
   e_{\mathrm{p}}(U)=\mathcal{F}_{d_1,d_2}-\mathcal{G}_{d_1,d_2}\{\mathrm{Tr}[(U^R{U^R}^\dag)^2]+\mathrm{Tr}[(U^{T_1}{U^{T_1}}^\dag)^2]\}.
\end{equation}
That means for the systems with certain dimensions, the entangling power is proportional to the trace of the product of the matrices after realignment and partial transposition to the time evolution operator. In other words, we have shown that the entangling power is completely determined by the realigned and partially transposed operators. These two matrix manipulations are sufficient to study the entangling capability of an unitary operator. The only task to get the entangling power is to calculate these trace terms.

For later uses, we provide a simple form of these trace terms. We can always decompose the evolution operator into the form:
\begin{equation}
\label{eq:decomposition_U}
  U=\sum\limits_{i}c_iA_i\otimes B_i.
\end{equation}
After some algebra, we get
\begin{equation}
  \label{eq:factor}
  \begin{split}
    \mathrm{Tr}[(U^R{U^R}^\dag)^2]= &\sum\limits_{ijkl}c_ic_j^*c_kc_l^* \mathrm{Tr}[A_i{A_j}^\dag]\mathrm{Tr}[A_k{A_l}^\dag]\mathrm{Tr}[B_i{B_l}^\dag]\mathrm{Tr}[B_k{B_j}^\dag],\\
    \mathrm{Tr}[(U^{T_1}{U^{T_1}}^\dag)^2]= &\sum\limits_{ijkl}c_ic_j^*c_kc_l^* \mathrm{Tr}[A_i{A_j}^\dag A_k{A_l}^\dag]\mathrm{Tr}[B_i{B_l}^\dag B_k{B_j}^\dag].
  \end{split}
\end{equation}
We can intuitively see the connection between the entangling power
and the decomposition of the evolution operator here. If the
evolution operator can be written as $U=U_1\otimes U_2$, we have
$\mathrm{Tr}[(U^R{U^R}^\dag)^2]=d_1^2d_2^2$,
$\mathrm{Tr}[(U^{T_1}{U^{T_1}}^\dag)^2]=d_1d_2$, then from
Eq.~(\ref{eq:ep_form}), the entangling power vanishes.
Equation~(\ref{eq:factor}) will be used when calculating the entangling
power of the Heisenberg interaction in Sec. IV.

\section{ Application I: Ising interaction}
\noindent
We study the entangling power of the time evolution of the Ising model, whose Hamiltonian reads $H=gS_1^z\otimes S_2^z$. The unitary time evolution operator generated by it is
\begin{equation}
  U=\sum\limits_{m_1=-s_1}^{s_1}\sum\limits_{m_2=-s_2}^{s_2}e^{i\theta m_1m_2} \ket{m_1m_2} \bra{m_1m_2},
\end{equation}
where $\theta = -g t$, $s_i$ is the length of spin $i$ and $\ket{m_i}$ is the eigenstate of $S_i^z$. $\hbar = 1$ is assumed for all cases.

Making use of the definition of realignment and partial transposition, we get
\begin{eqnarray}
  U^{T_1}&=&U,\\
  U^R&=&\sum\limits_{m_1m_2}e^{i\theta m_1m_2}\ket{m_1m_1} \bra{m_2m_2}.
\end{eqnarray}
Then
\begin{equation}
  \label{eq:delta}
  U^R(U^R)^\dag = \sum\limits_{m_1m_2}(\sum\limits_{n_1}e^{i\theta(m_1-n_1)m_2})\ket{m_1m_1} \bra{n_1n_1}.
\end{equation}
Note that the range of the indices in the above expression is all
symmetric, so we can substitute index $m$ with $-m$. Then
after trace, we get
\begin{equation}
  \label{eq:r_term}
  \begin{split}
    &\mathrm{Tr}[(U^R(U^R)^\dag)^2]\\
    &=\sum\limits_{m_1,n_1=-s_1}^{s_1}(\sum\limits_{m2=-s_2}^{s_2} e^{i\theta(m_1+n_1)m_2})^2\\
    &=d_1d_2^2+\sum\limits_{M=1}^{d_1-1}2(d_1-M)\frac{1-\cos(d_2 M \theta)}{1-\cos(M \theta)},
  \end{split}
\end{equation}
and
\begin{equation}
  \label{eq:p_term}
  \mathrm{Tr}[(U^{T_1}(U^{T_1})^\dag)^2]=\mathrm{Tr}[(UU^{\dag})^2]=d_1d_2,
\end{equation}
where $d_i=2s_i+1$ is the dimensions of the Hilbert space of spin $i$. Substituting Eqs.~(\ref{eq:r_term}) and (\ref{eq:p_term}) to (\ref{eq:ep_form}) leads to
\begin{equation}
  \label{eq:ep_ising}
    e_{\mathrm{p}}(U) = \frac{(d_1-1)d_2}{(d_1+1)(d_2+1)}-\frac{1}{d_1d_2(d_1+1)(d_2+1)}\sum\limits_{M=1}^{d_1-1}2(d_1-M)\frac{1-\cos(d_2 M \theta)}{1-\cos(M \theta)}.
\end{equation}
For the cases of the qubit-qudit system ($d_1=2$), we get
\begin{equation}
    e_{\mathrm{p}}(U)=\frac{d_2}{3(d_2+1)}-\frac{1-\cos(d_2\theta)}{3 d_2(d_2+1)(1-\cos \theta )}.
\end{equation}
And for the cases of two-qubit system ($d_1=d_2=2$), we get the same result as \cite{XWang1}
\begin{equation}
e_{\mathrm{p}}(U)=\frac{2}{9} \sin^2(\theta/2).
\end{equation}

\begin{figure}[htbp]
  \centering{
    \epsfig{scale=0.35, file=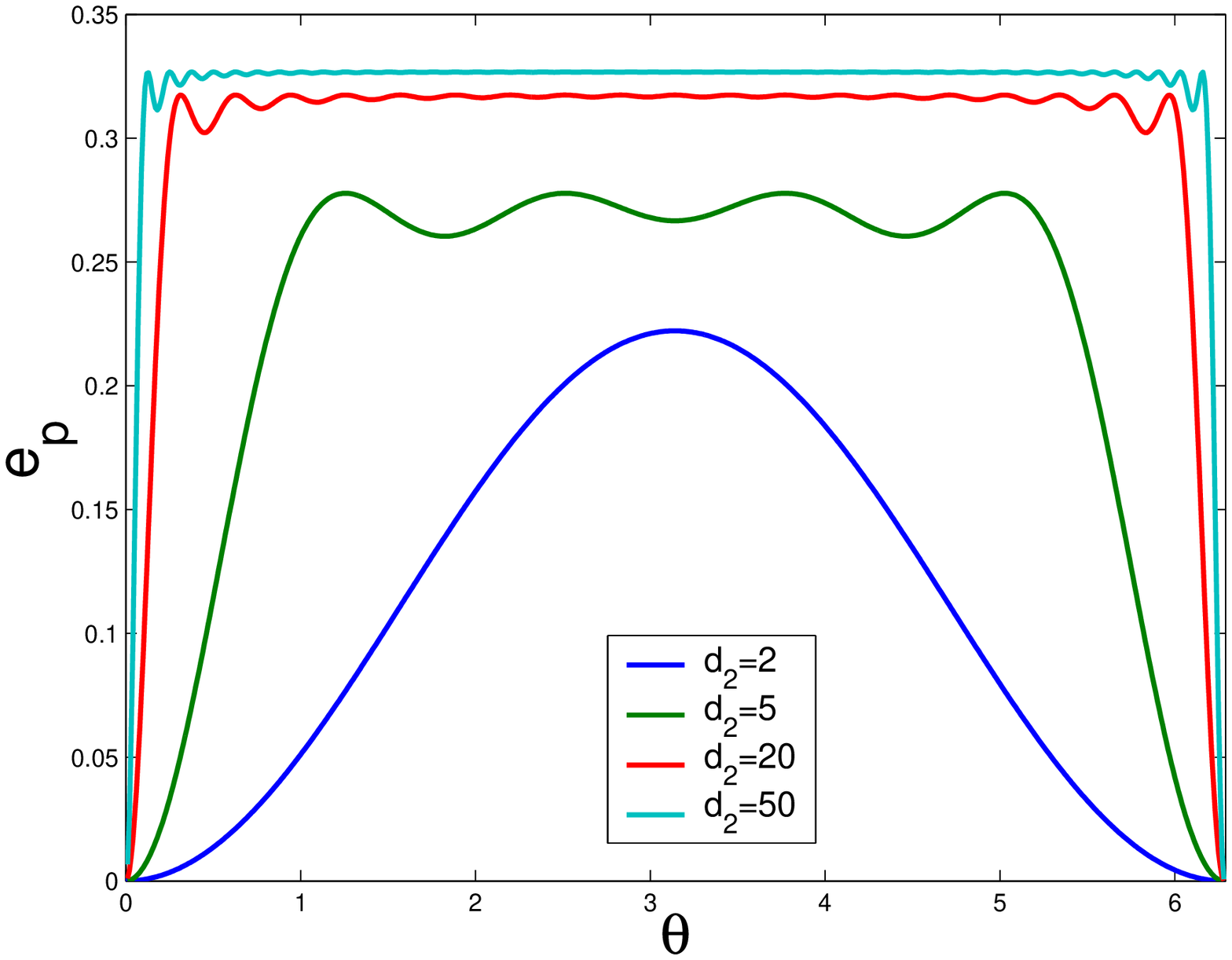}
    \epsfig{scale=0.35, file=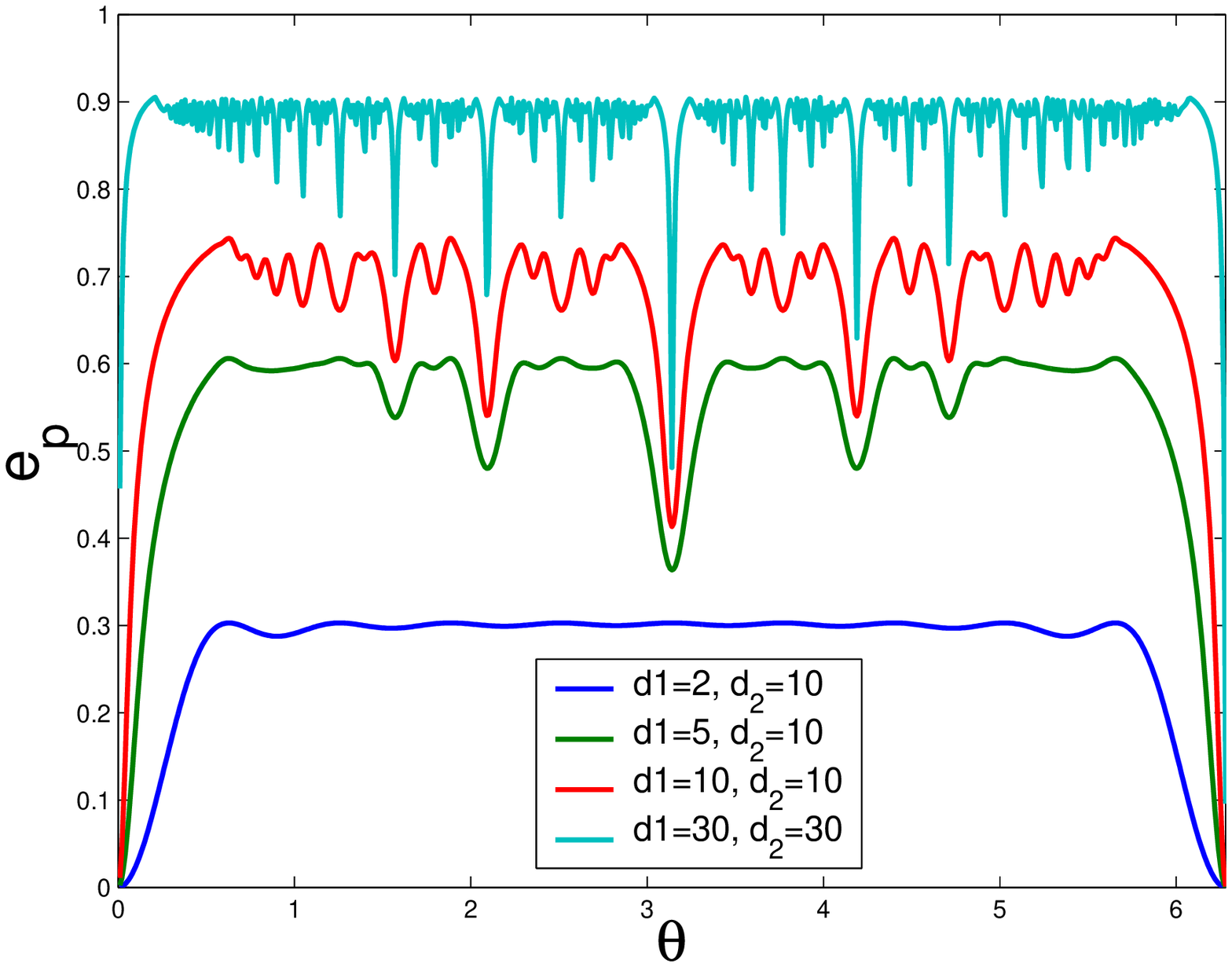}
  }
  \vspace*{13pt}
  \caption{\label{Fig:Ising}The entangling power of the Ising model. In the left figure we let $d_1=2$.}
\end{figure}

% \begin{figure}[htbp]
%   \centering{\epsfig{scale=0.5, file=epower_fig2.eps}}
%   \vspace*{13pt}
%   \fcaption{\label{Fig:Ising2}The entangling power of the Ising model, for different $d_1$ and $d_2$.}
% \end{figure}

From Fig.~(\ref{Fig:Ising}), we may see
that when the dimensions of the subsystems increase, the entangling
power increases on average, and the time oscillations become more
complex when dimensions increase. At $\theta=\pi$, the entangling
power reaches its extremum, but often it is a minimum other than
maximum. The maxima of entangling power mean large entanglement
generation, and these points correspond to some useful quantum
gates. So, the study of entangling power helps us to construct
quantum gates in quantum computation with hybrid qudits.

We further consider the time average of entangling power defined by Eq.~(\ref{eq:time_average_ep}). Because
\begin{eqnarray}
  \nonumber
  &&\frac{1}{2\pi}\int\limits_0^{2 \pi}\mathrm{Tr}[(U^R(U^R)^\dag)^2] d \theta\\
  \nonumber
  &=&\frac{1}{2\pi} \sum\limits_{m_1,n_1=-s_1}^{s_1}\sum\limits_{m_2,n_2=-s_2}^{s_2} \int\limits_0^{2 \pi}e^{i\theta(m_1+n_1)(m_2+n_2)}d\theta\\
  &=&d_1d_2(d_1+d_2-1),
\end{eqnarray}
then after substituting it to the definition of the time-average entangling power Eq.~(\ref{eq:time_average_ep}), we get
\begin{equation}
\overline{e_{\mathrm{p}}}(U)=(1-\frac{2}{d_1+1})(1-\frac{2}{d_2+1}).
\end{equation}
It increases as the dimensions of the subsystems increases, which can also be seen from Fig.~(\ref{Fig:Ising}) intuitively.

\section{ Application II: Heisenberg  interaction}
\noindent
We now consider a general SU(2)-invariant Hamiltonian. According to the Schur's lemma, it can be written as
\begin{equation}
  H=\sum\limits_{n=s_2-s_1}^{s_2+s_1} E_n P_n,
\end{equation}
where $P_n$ is projection operator of the total spin-$n$ subspaces and $E_n$ is the eigenenergy. Without loss of generality, $s_1\leq s_2$ is assumed here.
The time evolution operator generated by it is given by
\begin{equation}
  \label{eq:U_SU2}
  U=\sum\limits_{n=s_2-s_1}^{s_2+s_1}\alpha_nP_n,
\end{equation}
where $\alpha_n=e^{-i t E_n}$.

On the other hand, the spin projection operators can be expressed in term of the SU(2) operators $(\mathbf{S_1}\cdot \mathbf{S_2})^n$ as \cite{Batchelor,GMZhang}
\begin{equation}
  \label{eq:projection}
  P_n=\prod\limits_{\substack{k=s_2-s_1 \\ \neq n}}^{s_2+s_1}\frac{\mathbf{S_1}\cdot \mathbf{S_2}-\lambda _k}{\lambda_n-\lambda_k},
\end{equation}
where
\begin{equation}
  \lambda_k=\frac{1}{2}\left[ k(k+1)-s_1(s_1+1)-s_2(s_2+1)\right].
\end{equation}
For instance, in the case of $s_1=1/2$, we get
\begin{eqnarray}
  \label{eq:expr_proj}
  \nonumber
  P_{s_2-1/2}&=&\frac{d-1-4 \mathbf{S_1} \cdot \mathbf{S_2}}{2d},\\
  P_{s_2+1/2}&=&\frac{d+1+4 \mathbf{S_1} \cdot \mathbf{S_2}}{2d},
\end{eqnarray}
where $d=2s_2+1$ is the dimensions of the Hilbert space of the second spin.

Substituting Eq.~(\ref{eq:projection}) into (\ref{eq:U_SU2}), we formally get the following useful expression
\begin{equation}
  U=\sum\limits_{n=0}^{2s_1}\beta_n(\mathbf{S_1}\cdot \mathbf{S_2})^n,
\end{equation}
where $\beta_n$ is the corresponding parameter. Substituting $\mathbf{S_1}\cdot \mathbf{S_2}=\sum_{i=x,y,z}S_1^i \otimes S_2^i$ into the above expression and expand it, then we get the decomposition form as in Eq.~(\ref{eq:decomposition_U}).

For an isotropic Heisenberg interaction bipartite model, the Hamiltonian reads $H=\mathbf{S_1}\cdot \mathbf{S_2}$. In the case of $s_1=1/2$, due to Eq.~(\ref{eq:expr_proj}), we can write the Hamiltonian as
\begin{equation}
  H=E_0P_{s_2-1/2}+E_1P_{s_2+1/2},
\end{equation}
where $E_0=-(d+1)/4$, $E_1=(d-1)/4$ are the eigenvalues of the Hamiltonian. And we can write the time evolution operator as follows
\begin{eqnarray}
  \nonumber
  U&=&\beta_0 \textrm{I} + \beta_1 \mathbf{S_1}\cdot \mathbf{S_2},\\
  \beta_0&=&\left[(d-1)e^{-itE_0}+(d+1)e^{-itE_1}\right]/(2d),\\
  \nonumber
  \beta_1&=&2(e^{-itE_1}-e^{-itE_0})/d,
\end{eqnarray}
where $\textrm{I}$ is the identity operator. Making use of Eq.~(\ref{eq:factor}), and noticing that
\begin{eqnarray}
  \textrm{Tr}(S^i S^j)&=& \delta _{ij}\sum\limits_{k=-s}^{s}k^2=\delta_{ij}\frac{1}{12}(d^2-1)d, \\
  \textrm{Tr}(S^i) &=& 0,
\end{eqnarray}
where $i \in \{ x,y,z \} $, we get
\begin{equation}
\mathrm{Tr}[(U^R{U^R}^\dag)^2]= 4d^2|\beta_0|^4+\frac{1}{192}(d^2-1)^2d^2|\beta_1|^4.
\end{equation}

On the other hand, for the partial transpose operation, we can use the partial time reversal \cite{Breuer} instead, which is defined by
\begin{equation}
  \label{eq:partial_time_reversal}
  M^{\tau_1}=e^{-i\pi S_1^y}M^{T_1}e^{i\pi S_1^y}.
\end{equation}
We have
\begin{equation}
  \textrm{Tr}((U^{T_1}{U^{T_1}}^\dag)^2)=  \textrm{Tr}((U^{\tau_1}{U^{\tau_1}}^\dag)^2).
\end{equation}
Then we get
\begin{eqnarray}
  U^{\tau_1}&=&\beta_0 I-\beta_1 \mathbf{S_1}\cdot \mathbf{S_2}=\gamma_0 P_0+\gamma_1 P_1,\\
  \nonumber
  \gamma_0&=&\beta_0-E_0\beta_1,\quad \gamma_1=\beta_0-E_1\beta_1,
\end{eqnarray}
where $P_0=P_{s_2-1/2}$ and $P_1=P_{s_2+1/2}$ are the projection operators of the subspaces with the eigenenergy $E_0$ and $E_1$, respectively. Hence we obtain
\begin{equation}
  \mathrm{Tr}[(U^{T_1}{U^{T_1}}^\dag)^2]= (d-1)|\gamma_0|^4+(d+1)|\gamma_1|^4.
\end{equation}
At last, we get the entangling power
\begin{equation}
e_{\mathrm{p}}(U)=\frac{4(d-1)}{9d^4}[3d^3-f(d)\sin^2(\frac{dt}{2})]\sin^2(\frac{dt}{4}),
\end{equation}
where $f(d)= 2(6-d+d^3)$.
In the special case of $d=2$, which means a two-qubit system, the above equation reduces to
\begin{equation}
  e_{\mathrm{p}}(U)= \frac{1}{6}\sin^2(t),
\end{equation}
which is consistent with the entangling power of a SWAP operator $S_{1,2}=2\mathbf{S_1}\cdot\mathbf{S_2}+1/2 $ \cite{ZMa}.
And the time-average entangling power is
\begin{equation}
  \overline{e_{\mathrm{p}}}(U)=\frac{(d-1)(d^3+d-6)}{3d^4}.
\end{equation}
It increases when the dimensions of the subsystems increases, as the case of the Ising interaction.

\begin{figure}[htbp]
  \centering{\epsfig{scale=0.4, file=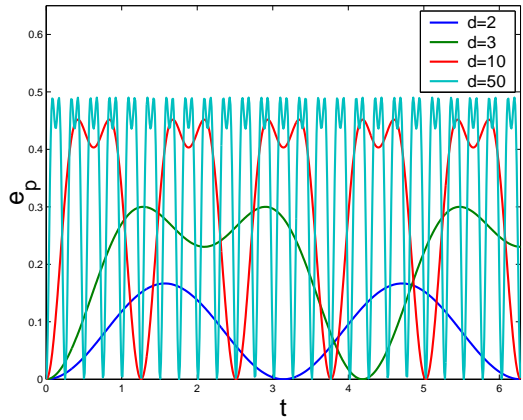}}
  \vspace*{13pt}
  \caption{\label{Fig:Heisenberg}The entangling power of the isotropic Heisenberg model.}
\end{figure}

We also make numerical calculations which are shown in
Fig.~(\ref{Fig:Heisenberg}). We can see that the period of the
entangling power is smaller when $d$ is greater. This is due to the
propertis of the time evolution operator. Because of the
SU(2) invariance, the Hamiltonian describing the isotropic
Heisenberg interaction between a qubit and a qudit system has only
two different eigenvalues, and the energy gap becomes greater as the
dimensions of the Hilbert space of the qudit subsystem increases,
then the period of the time evolution becomes smaller.
\section{Conclusion}
\noindent
In conclusion, we have extended the matrix rearrangement approach for the entangling power to the general cases of the hybrid qudit($d_1 \times d_2$) systems. This approach supplies a convenient way to get the entangling power, and we show that the exact solutions of the entangling power are obtained for some simple Hamiltonians. We also consider the effects of the dimensions of the subsystems, and find that the time average entangling power are monotone increasing functions with respect to the dimensions of the subsystems.

Comparing the formula given by Zanardi et al.~\cite{Zanardi}, the
present one is simpler to apply. What we need to do is only to
calculate the realigned matrix and the partially transposed matrix,
and after making some simple traces, we can obtain the entangling
power. The entangling power has been successfully used in the study
of quantum chaos~\cite{Scott}, and we believe that the time-average
entangling power introduced here is also useful for characterizing
nonlinear behaviors of quantum systems.

\section{Acknowledgements}
\noindent
We thank Z. Ma for helpful discussions. This work is supported by NSFC with grant Nos. 10405019 and 90503003; NFRPC with grant No. 2006CB921205; Program for new century
excellent talents in university (NCET). Specialized Research Fund for the
Doctoral Program of Higher Education (SRFDP) with grant No. 20050335087.

\end{document}